\documentclass[12pt]{iopart}

\usepackage{iopams}  
\usepackage{graphicx}  
\usepackage{color}  
\begin{document}

\title{Simple interferometric setup enabling sub-Fourier-scale ultra-short laser pulses}

\author{Enrique G. Neyra, Gustavo A. Torchia, Pablo Vaveliuk and Fabian Videla$^{1,2}$}

\address{$^1$ 
Centro de Investigaciones \'Opticas (CICBA-CONICET-UNLP), Cno. Parque Centenario y 506, P.O. Box 3, 1897 Gonnet, Argentina}
\address{$^2$
Departamento de Ciencias B\'asicas, Facultad de Ingenier\'ia UNLP, 1 y 47 La Plata, Argentina}

%
\ead{enriquen@ciop.unlp.edu.ar}

\date{\today}




\begin{abstract}

In this work,  we describe an interferometric method to generate ultra-short pulses below the Fourier limit. It is done by extending concepts first developed in the spatial domain to achieve sub-diffractive beams through the addition of a spatial chirp in one of the arms of a Michelson interferometer using a spherical mirror. To experimentally  synthesize sub-Fourier pulses, we replace the spherical mirror with a water cell, since it produces chirp in the  temporal domain.
We also present an alternative procedure, based on asymmetrical interference between the widened pulse and the original pulse where the peaks of both pulses  exhibit a temporal delay achieving the narrowing of  ultra-short pulses with  sub-Fourier scales.

To characterize the performance of the system, we  performed  a preliminary assessment considering the percentage of FWHM shrinking obtained for each scheme. By means of a symmetrical configuration 7 and 12 \%  pulse reductions were verified, both experimentally and analytically, while for the non-symmetrical configuration 10 and 24\% reductions were achieved corresponding to main lobe to side-lobes ratios of 10 and 30\% .

The experimental setup scheme is simple, versatile and  able to work with high-power laser sources and  ultra-short pulses with a broad bandwidth at any central wavelength. The results presented in this work are promising and  help to enlighten new routes and strategies in the design of coherent control systems. We envision that they will become broadly useful in different areas from strong field domain to quantum information.

\end{abstract}

%
%
%
%
%


\section{Introduction}

The  temporal  width  of   an  ultra-short  laser pulse $E(t)$, is given by the distribution  of the  spectral phase $\phi(\omega)$ in  the Fourier spectral components of the field $E(\omega)$. The so-called  transform-limited pulses condition is fulfilled if the spectral phase of the field is independent of the  frequency. This condition implies that the temporal and spectral widths are minimum and in turn, its product corresponds to a particular constant in agreement with the  pulses shapes. Considering a Gaussian pulse, and their temporal and spectral full width at half-maximum (FWHM) , this product yields : $\Delta\omega$ $\Delta t = 2 \pi c $ with $c=0.441$ \cite{diels2006ultrashort}. Of course, the c value  changes for other pulse shapes.

The Fourier limit is tied to the quantum Heisenberg uncertainty principle since the conjugate variables of both phenomena are related by the Fourier transform. In contrast, if the pulse does not meet this condition, i.e. the spectral phase $\phi(\omega)$ is frequency dependent, the pulse presents  \textit{chirp} and the FWHM of the pulse is larger than the FWHM of the pulse without a \textit{chirp}.

 In this sense, there  has been an emerging interest in recent years for a  wave phenomenon called superoscillation
 \cite{aharonov2011some,berry2019roadmap}, in which a band-limited function can oscillate faster than their fastest Fourier component.

The superoscillatory phenomenon can be seen as an interferometric one, where each spectral component has a particular phase that originate, in the conjugate space, small regions whose FWHMs are below the Fourier limit. Additionally, those regions where the Fourier limit is "broken", show poor efficiency about the initial pulse. In addition, this phenomenon coexists with the emerging of side-lobes with relative amplitude \cite{neyra2021tailoring1}. 



The superoscillatory phenomenon is not only an interesting  mathematical topic. Several wave phenomena take advantage of superoscillations such as: Focusing different kinds of beams for superresolution microscopy  \cite{gbur2019using}, in quantum mechanics \cite{remez2017superoscillating,berry2006evolution}, in acoustic waves \cite{shen2019ultrasonic}, in signals \cite{zarkovsky2020transmission}, and others\cite{berry2019roadmap}. Additionally, for ultra-short pulses, the Fourier limit can be overcome through devices like the  spatial light modulator(SLM) in a prism based pulse shaper  \cite{weiner2000femtosecond}, which generates the desired phase mask \cite{boyko2005temporal,binhammer2006spectral,eliezer2018experimental,eliezer2017breaking,rausch2008few}. However, this experimental setup requires a complex process of alignment and control,  complemented  with cumbersome numerical tools. On the other hand, limitations in the beam shrinking process using pulse shapers (SLM's), arise specially from the chromatics aberrations because of the bandwidth of short pulses. On the other hand pulse shapers need of phase calibration. Likewise, must be mentioned that SLM's presents aberrations given by their pixel size \cite{eliezer2018experimental}.


It is an accepted notion that the minimum  reachable FWHM  is associated with a constant spectral phase, but this is not strictly true. 
The clearest example is put forward by a $\pi$-phase mask where sub-Fourier pulses were obtained  \cite{eliezer2018experimental}. Additionally, the ability to manipulate the \textit{shape} of the optical pulses is essential for numerous applications  \cite{keller2003recent} such as spectroscopy and coherent control \cite{goswami2003optical,silberberg2009quantum}, metrology \cite{tanabe2002spatiotemporal} , microscopy \cite{bardeen1999effect} and optical communications \cite{sardesai1998femtosecond}.

In this article,  we propose a robust, versatile, and simple method to achieve sub-Fourier ultra-short laser pulses.
The  current  approach is inspired by the spatial narrowing of a focused beam, recently proposed \cite{neyra2021tailoring1}. In that work, to print a quadratic phase in the beam spatial profile, a spherical mirror was introduced modifying  a branch of a Michelson interferometer.  
In this work to achieve a quadratic phase in the beam but now in the temporal domain, we replaced the spherical mirror with a water cell (dispersive media) introducing  \textit{chirp} on the original pulse,  As a result,  a temporally widened pulse is obtained after the cell. To assess the performance of the proposed experimental scheme, we developed two experiments.
The first one is based on a symmetrical interference between the widened pulse and the original one, that is, when the peaks of both pulses coincide. The second one, is based in an  asymmetrical interference between the widened pulse and the original pulse, where the peaks of both pulses exhibit a temporal delay.

\section{Theoretical framework}

In order to mathematically describe the interfering process, we call $E(t)$ the pulse originated as a result of the superposition of
\textit{pulse 1} (unaltered pulse) and \textit{pulse 2} (altered pulse) after passing  through the interferometric device. \textit{Pulse 1} is a band-limited ultra-short Gaussian pulse $E_0(t)$, with spectral content $E_0(\omega) = E_0 e^{-((\omega-\omega_0)/\delta\omega)^2}$, centered at the carrier frequency $\omega_0 $ whose  bandwidth is represented by  $\delta\omega$.
Therefore, the whole field in the frequency domain, composed by the unaltered and the altered pulse is given by:

%
\begin{equation}
E(\omega)=E_0(e^{-(\frac{\omega-\omega_0}{\delta\omega})^2}+\alpha e^{-(\frac{\omega-\omega_0}{\delta\omega})^2}e^{-i \beta (\omega-\omega_0)^2}e^{i \phi})\label{e2}
\end{equation}


Clearly, the phase of \textit{pulse 2} is modified by the chirp parameter $(\beta)$, in turn the amplitude is  modified by  $(\alpha)$, a factor that weighs the relative amplitude between the arms. Besides, there is a third controllable parameter, $(\phi)$,  representing the relative phase between both arms of the interferometer. The Fourier transform of the Eq. (\ref{e2}) returns the narrowed pulse in the time domain: 
\begin{equation}
E(t)=e^{-\frac{1}{4}t^2}e^{i \omega_0 t}+\frac{\alpha}{(1+\beta^2)^{1/4}} e^{-\frac{1}{4(1+\beta^2)}t^2} e^{i \Phi(t)}e^{i \omega_0 t}, \label{e3}
\end{equation}
where the $t$-varying phase is:
\begin{equation}
\Phi(t)=\frac{\beta}{4(1+\beta^2)}t^2-\frac{1}{2}\arctan{(\beta)}+\phi\label{e4}.
\end{equation}
The Fourier transform was evaluated taken $\delta\omega$ and $E_0$ equal to the unit, for a sake of clarity to visualize the technique highlights.


Explicitly, Eq. (\ref{e3}) shows the interference between a Fourier-transform-limited pulse $E_0(t) = e^{-\frac{1}{4}t^2}$ and the widened version of this pulse, broadened by the factor $\tau =\sqrt{ 1+\beta^2}$ with relative  amplitude $\alpha/(1+\beta^2)^{1/4}$ and phase $\Phi(t)$.

It is well-known that the chirp quadratic parameter $\beta$, in the temporal domain, is originated when the pulse travels through  a dispersive medium. This parameter is related to the group velocity dispersion (GVD) by  $\beta=\frac{1}{2}GVD\times L$, where $L$ is the length of the dispersive media \cite{diels2006ultrashort}.  %
The pulse intensity $I(t)$ is then obtained by multiplying Eq.\ref{e3} by its complex conjugate to give:
\begin{equation}
I(t)=e^{-\frac{1}{2} t^2}+\frac{\alpha ^2 }{\sqrt{1+\beta ^2}}
e^{-\frac{t^2}{2(1+\beta ^2)}}+\frac{2\alpha }{(1+\beta ^2)^{1/4}}e^{-\frac{t^2}{4}(1+\frac{1}{(1+\beta^2)})}\cos{\Phi}\label{e5}
\end{equation}

Equation (\ref{e5}) represents the temporal intensity profile of the field $E(t)$, that for a suitable choice of the external parameters $\beta$, $\alpha$ and $\phi$, make it possible to shape a pulse whose central lobe shows a temporal width below the Fourier limit, i.e. narrower than the reference pulse $E_0(t)$. 
For this purpose, a pure destructive interference condition between the pulses was necessary,
which was achieved for values of the phase $\phi=\phi(\alpha,\beta)$, $\alpha$ and $\beta$. A complete theoretical description of this interference was made in Ref \cite{neyra2021tailoring1}.
There, in the same plot,  the reduction of the FWHM are represented as a function of $\alpha$ and $\beta$, as well as, curves bounding regions for different ranges of the ratio between the side-lobes and the central lobe. Superimposed on the previous regions  a third region  with a high efficiency for the central lobe, a trade-off  between the previous zones, making the performance of the technique suitable for each application.


\section{Experimental}

\subsection{Setup}

The experimental setup used to carry out the proposed approach to synthesize narrower pulses is sketched in Figure 1.  The ultra-short pulse input into a Michelson Interferometer (MI), then  was split into two pulses by means of a beam-splitter (BS). \textit{Pulse 1} traveled through branch 1 of the interferometer and then was reflected by the fixed mirror (M1), remaining unaltered. 
The \textit{pulse 2} traveled through branch 2, this arm undergoing the influence of the different control parameters, $\alpha$, $\beta$ and $\phi$. Mirror (M2) was mounted in a micro-meter translation stage and controls the relative phase $\phi$ between the interferometer arms. During the first experiment, MI was set as a balanced interferometer, while for the second a delay line $\tau_D$ was activated  to accomplish a non-symmetrical interference.
In branch 2,  \textit{pulse 2} was stretched by passing it through a 6 cm long water cell, due to the group velocity dispersion ($\beta$ parameter, GVD).The path through the cell (indicated as E) is shown in  Fig.1.  In the same branch, \textit{pulse 2} passes through a half-wave plate (HWP) and a polarizer (P) system to control the relative intensity between both pulses ($\alpha$ parameter), device D in Fig. 1. 
Finally, when the two pulses are reflected by the mirrors ( the pulse cross again the water cell), the resulting interference is detected by a GRENOUILLE system \cite{akturk2003measuring}. GRENOUILLE can be considered as an implementation of  FROG (Frequency Resolved Optical Gating) but without mobile parts, and was indicated by
F in the setup. Its function was to characterise the temporal profile of the pulse synthesis.
The laser system employed to test our experimental approach was a Ti:Sapphire femtosecond oscillator (S), Mai Tai model from Spectra Physics (USA), characterised by  a spectral bandwidth of 10:5nm, with a band-limited pulse of FWHM $\approx$ 90fs (see inset of Fig. 1).
\begin{figure}[hb]
 \centering
\includegraphics[width=0.7\textwidth]{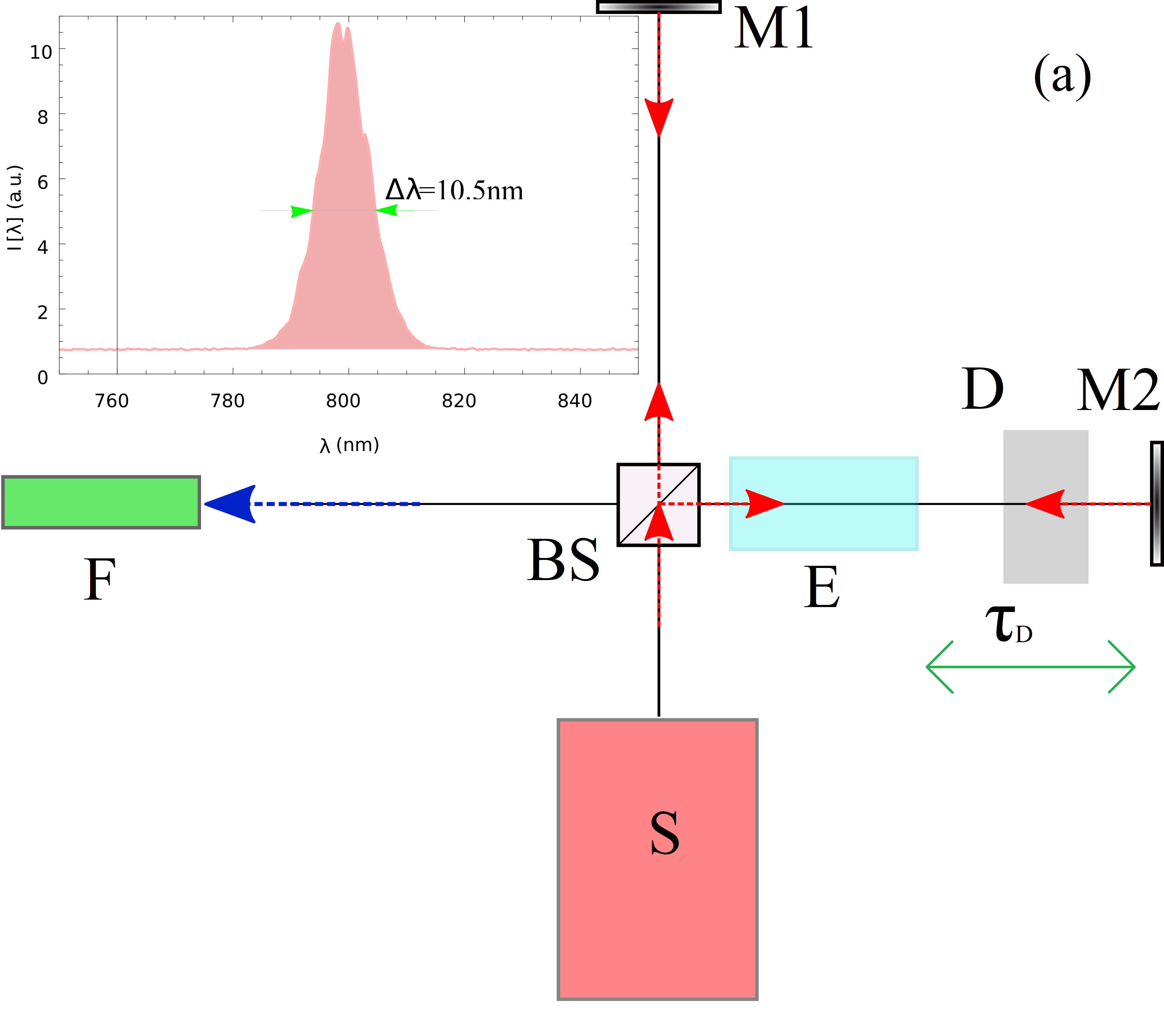} 
\caption{Our experimental setup for synthesise pulses is based on a Michelson Interferometer (MI). 
S is the femtosecond laser source, a Ti:Sapphire femtosecond oscillator. M1 and M2 are the fixed and mobile mirrors respectively. BS is the beam splitter. 
In the delay line, the water cell E (dispersive medium) 
is followed by a half-wave plate (HWP) plus a polarizer, D, 
at the opposite end, the GRENOUILLE system F. It was indicated  the movable part in the arm for $\tau_D$ control
\label{Fg}}
\end{figure}

\subsection{Results and discussions}


First of all, we measured the different pulses widths,  which are represented in Figure 2. Panel (a) shows the FROG trace for the pulse into branch 1, while Fig 2(b) shows the temporal intensity  profile and its  phase retrieved by  the FROG trace, with temporal width of FWHM=89.8fs. Panel (c) of Figure 2 shows the measured FROG trace corresponding to the stretched pulse by propagation through the water cell (branch 2). Panel (d) presents the FROG retrieved temporal intensity and phase profile of the pulse,on which an FWHM can be observed of 199,4fs. By properly setting  units and considering  $\delta\omega=1$  the relationship between the FWHMs of the pulses obtained in each branch was $\sqrt{1+\beta^2}$ (see Eq.\ref{e3}). In this way the  $\beta$ value could be solved as $\frac{199.4fs}{89.8fs}=\sqrt{1+\beta^2}$ which results in $\beta=1.99$ (in fractions of $\delta\omega$). 

\begin{figure}[hb]
 \centering
\includegraphics[width=0.7\textwidth]{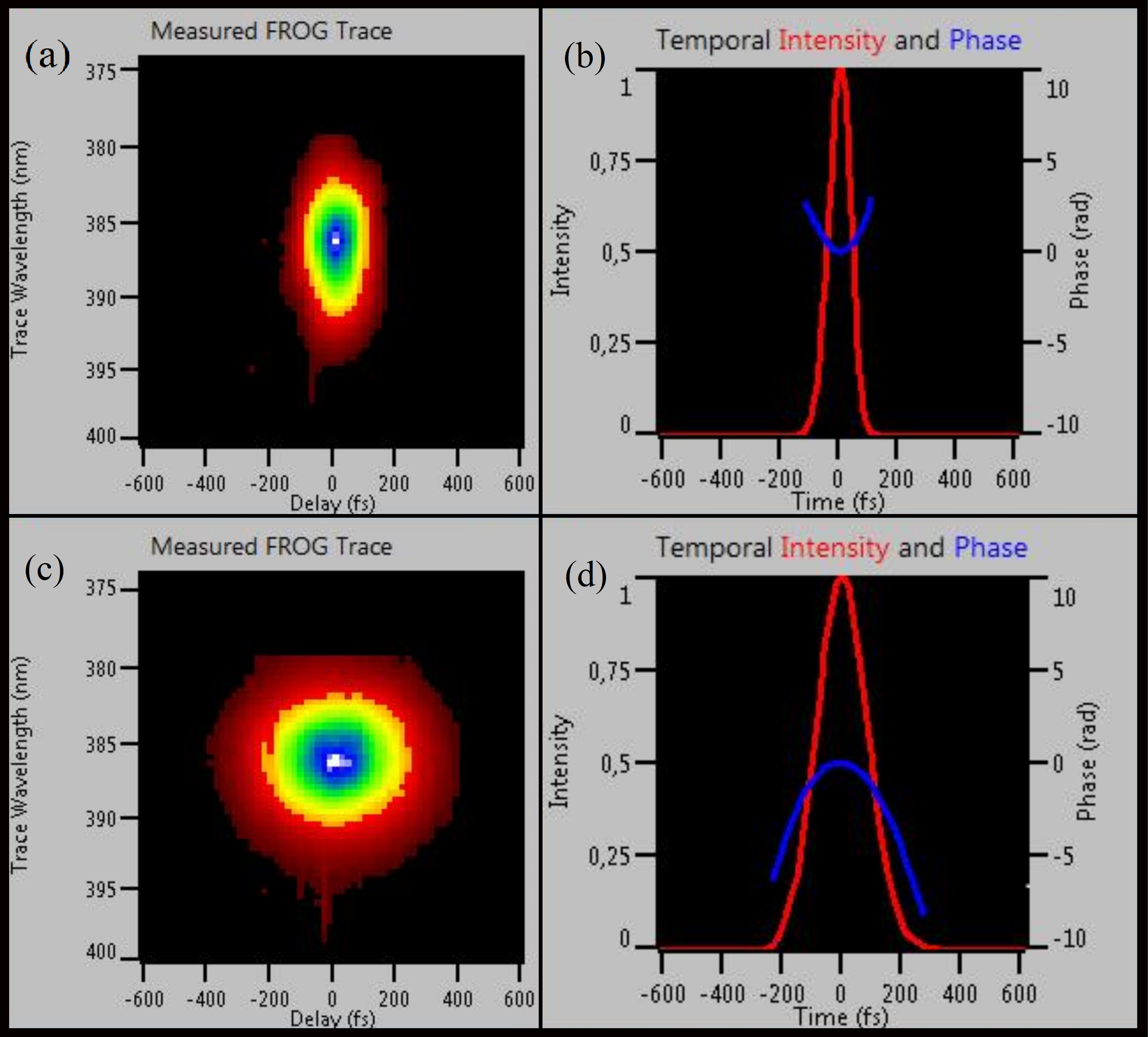} 
\caption{FROG traces corresponding to  pulses traveling in each arm of the Michelson Interferometer (left side). Retrieved temporal profiles for the pulses in each arm of the MI. (right side). \label{fig2}}
\end{figure}


Fig 3 presents the synthesized pulses  obtained  with the interferometric technique  described previously. The different pulses widths showed in panels (b) and (d) were obtained by varying the $\alpha$ parameter handling a  half-wave plate plus a polariser system, detailed in the experimental procedure. 
In Figure 3, panels (a) and (b) show the FROG trace and its retrieved pulse profile, respectively. The result of the interference was measured by setting $\alpha=0.25$. The FWHM retrieved was 84 fs.  corresponding  to a reduction of the Fourier limit of 7 percent for the input pulse. 
On the other hand, in Figures 3(c) and 3(d), the second set of measurements is shown. In this case, the interference pulse was conducted by setting $\alpha=0.38$. The retrieved FWHM was 79 fs,  corresponding to a reduction of 12\% of the original pulse limited by Fourier transform. 

\begin{figure}[hb]
 \centering
\includegraphics[width=0.7\textwidth]{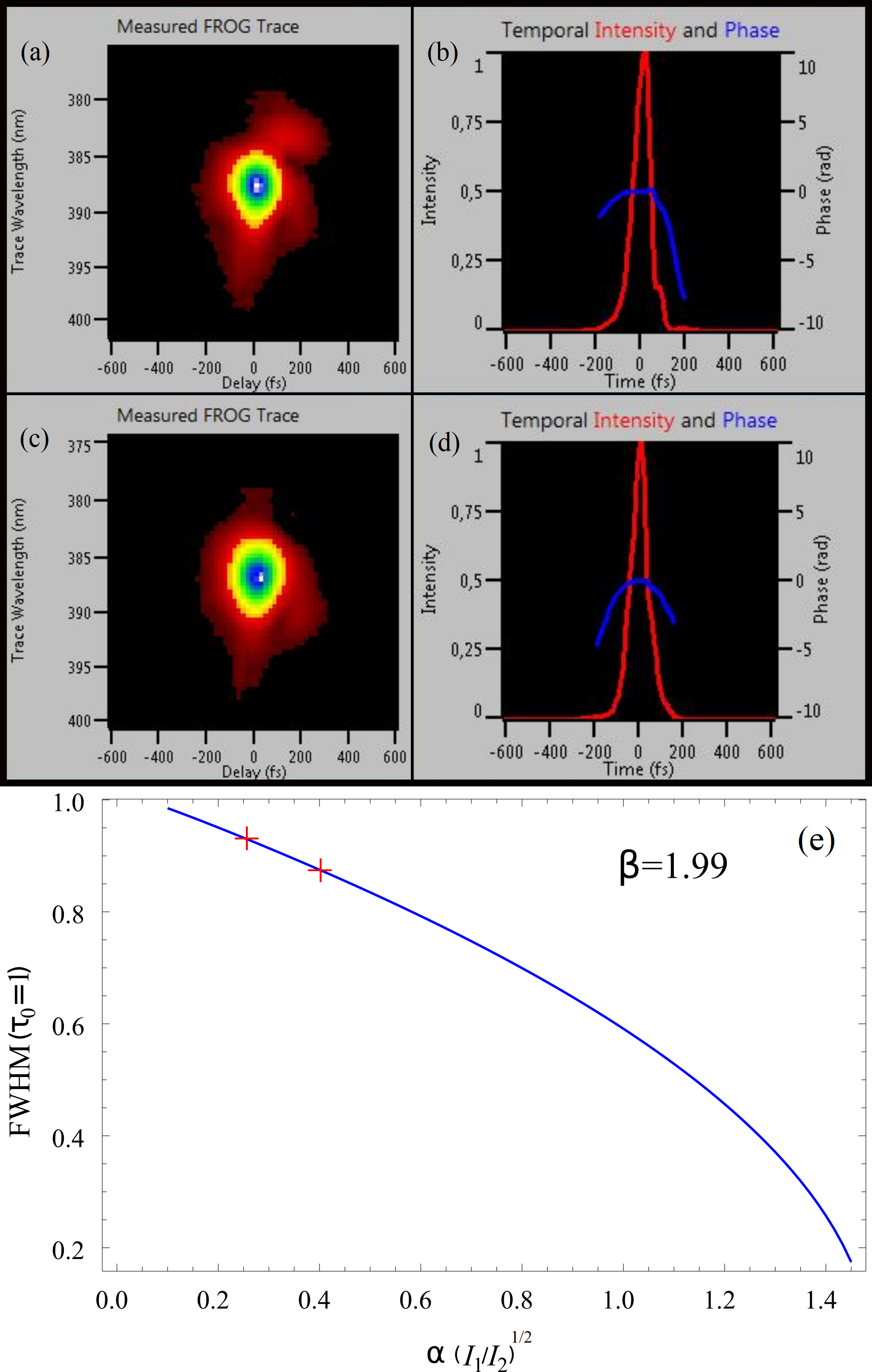} 
\caption{FROG traces corresponding to synthesised pulses with parameters $\alpha=0.25$, $\beta=1.99$ and $\alpha=0.38$, $\beta=1.99$ are sketched in  panels (a) and (c), respectively. In panels (b) and (d) the temporal profile of the Frog traces (a) and (c), where it can be seen a FWHM of 84 fs and 79 fs respectively. In (e), evolution of FWHM of the synthesised pulses  vs $\alpha$ (the amplitude ratio of the interfering pulses),  in fractions of the unaltered FWHM pulse considered as unit, so $\tau_0=1$. The chirp parameter $\beta$ was held constant at 1.99. The two red crosses correspond to the measured experimental data.\label{fig3}}
\end{figure}

In a complementary fashion, Fig 3 (e) shows the evolution of the FWHM corresponding to the central lobe for the synthesised pulse as a function of the $\alpha$ parameter (solid line). The FWHM scale was graduated in reference  to the temporal FWHM of the input pulse. This simulation, described by equation 4, was made considering a fixed value of the  $\beta$ parameter equal to 1.99. The first region from $\alpha$=1 to $\alpha$=0.8, shows a linear relation between FWHM and $\alpha$. A second region, for $\alpha > 1 $, presents a fast reduction of the FWHM.  By calculation of equation (4), for large values of $\alpha$, the synthesised pulse shows side-lobes with larger amplitude compared to the central lobe, as well as when the FWHM is close to zero value. As  described in \cite{neyra2018synthesis}, each side-lobe has a different spectral content and a complete theoretical analysis can be found in Ref. \cite{neyra2021tailoring1}. 
In Figure 3, experimental points have been indicated by red crosses. In this case, it is important to mention that we could not take measures for large values of $\alpha$, to verify the theoretical curve (Fig 3(e)), because of the technical limitations  of the GRENOUILLE system. When measures were made with large values of $\alpha$, interference rings are generated in the wave-front of the synthesised pulse. This could be possibly explained by the additional divergence of the pulse caused when passing through the water cell. It is necessary to work with Gaussian or continuous  wave-fronts to achieve a proper operation of the  GRENOUILLE system \cite{akturk2003measuring}.

Alternatively to the above method, we propose another way to produce similar sub-Fourier synthesized pulses controlling the delay times $\tau_D$ and the attenuation parameter $\alpha$ by using the same Michelson interferometer. In this case, the method is based on a non-symmetrical interference in the time domain. As we show, Eq. 1 represents the interference between an ultra-short and a widened pulse centred in $t=0$, which produce a synthesised pulse with sub-Fourier characteristics. The proposed alternative can be achieved if the two pulses traveling in each arm of the MI, the original and widened pulse, have a delay time $\tau_D $ at the moment of the interference. We will show that changes in the delay of the interferometer branches produce similar behavior that only changes the $\alpha$ parameter. Both lead to  sub-Fourier pulses. 


In Fig. 4  two measurements corresponding to different delay times ($\tau_D$) are presented (see Fig. 1), considered as the distance between peaks of the original and widened pulses, respectively (see fig(4)). The chosen values are :  $\tau_D =31.5$fs($\alpha=0.85$) and $\tau_D=49.5$fs ($\alpha=0.5$), top and bottom figures, respectively. In panels 4(a) and 4(b) at the top of Fig. 4 show the FROG trace and the retrieved temporal profile corresponding to a delay time of $31.5$fs  and $\alpha=0.85$; the synthesised pulse retrieved had an FWHM=68fs ($24\%$ of the Fourier limit reduction). In the  middle  of Fig 4. panels 4(c) and 4(d) respectively  show the FROG trace and the retrieved temporal profile corresponding to a delay of $49.5$fs and $\alpha=0.5$, giving a resultant pulse with an FWHM=81fs ($10\%$ Fourier limit reduction). Note that $\alpha$ is the relation between the maximum amplitude of both fields. When different delays are introduced the peak amplitudes are shifted, resulting in an effective amplitude ratio between pulses smaller than $\alpha$, since the interference  occurs while the peak of the unaltered pulse overlaps with the tail of the widened pulse. At the bottom of the  figure, panels 4(e) and 4 (f) show the respective simulations of the measurements. 
The blue line designates the original pulse and in green the widened pulse, with the delay $\tau_D$. 
The synthesised pulse after the interference is shown by the  red dashed line. Here, the asymmetric distribution of the  two side-lobes can be observed and are in good agreement with  the experimental measurements.

This last alternative to obtain sub-Fourier pulses allows modulating the temporal width of the pulse as a function of the temporal delay $\tau_D$. This process of FWHM modulation can be described through the next sequence:  suppose we began with two pulses temporally spaced by a large delay $\tau_D$,  which is progressively reduced through the movement of the micro-positioning stage. After that, the Fourier limited pulse will experience a reduction of FWHM. The minimum FWHM is reached when a maximum of the destructive interference condition occurs. If $\tau_D$ is reduced yet more, the Fourier limited pulse will begin to increase again. 
Depending on the FWHM of the widened pulse, this process is repeated once, twice or $n$ times.
In consequence, the temporal width of the initial pulse, will experience a sinusoidal variation with the  $\tau_D$ parameter. 

For a better understanding and visualisation of the previous description, two simulations were uploaded [ref] as supplementary material. Each of them represents, qualitatively, the change of the FWHM when the delay  $\tau_D$ varies. Both set the $\alpha$ parameter as one while  the  $\beta$  parameter as 5 and 10, respectively. Simulations, carefully observed, shows an FHWM oscillatory behaviour.  

Finally, it is worth mentioning that the experimental results presented in this work are limited by our detection system (GRENOUILLE). We think that a complete set of measurements,  predicted theoretically in Ref. \cite{neyra2021tailoring1}, can be achieved by another detecting system like FROG or SPIDER, as other authors have shown \cite{boyko2005temporal,eliezer2017breaking} since in the conjugate space, the different techniques are similar. In our case, the phase mask necessary to obtain sub-Fourier pulses is given by the operator $H(\omega)=(1+\alpha e^{i\beta( \omega - \omega_0)}e^{i\phi})$ (see Eq.\ref{e2}), either  implemented by a SLM or an interferometric system as in our case.

\begin{figure}[hb]
 \centering
\includegraphics[width=0.7\textwidth]{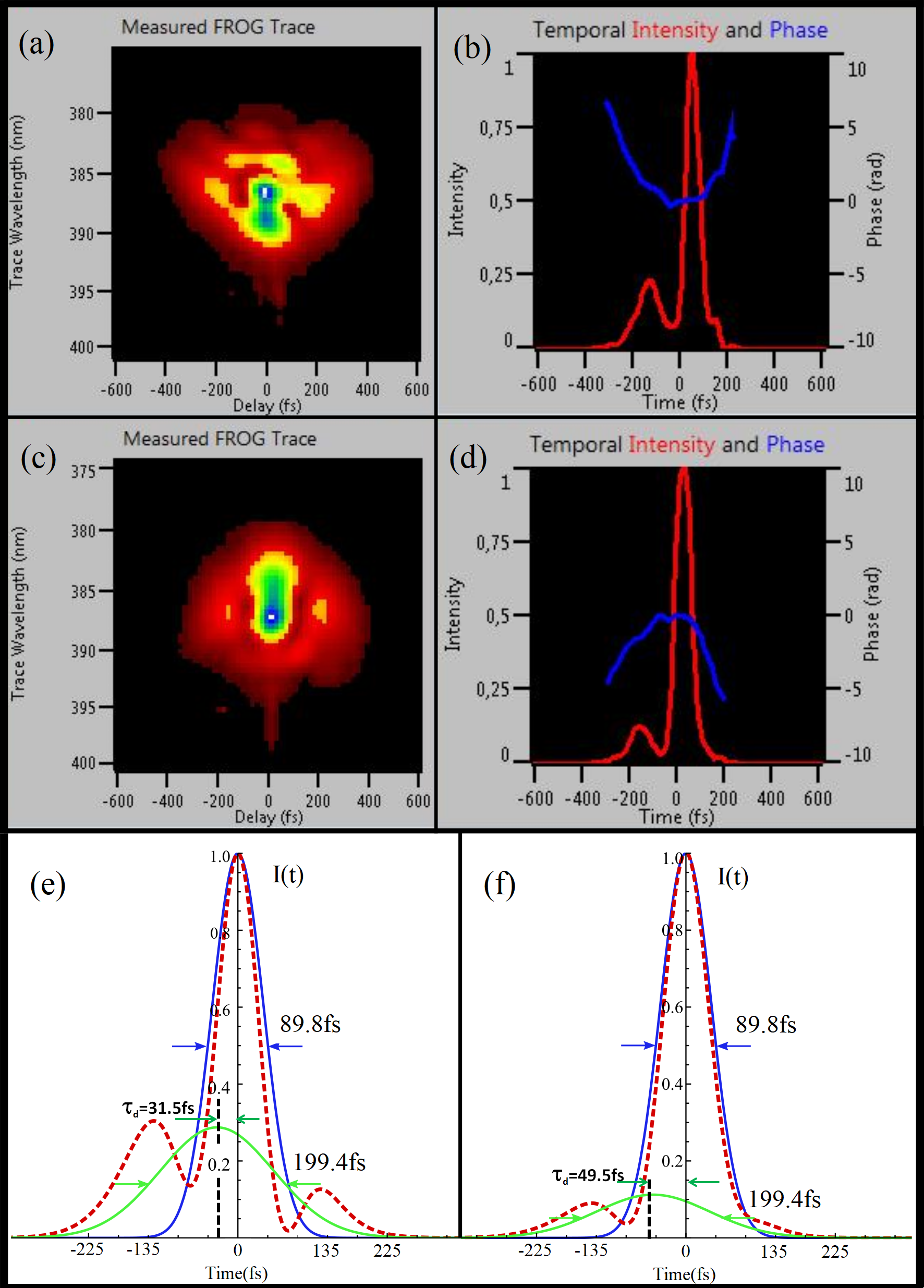} 
\caption{FROG traces corresponding to synthesised pulses for delays of 31.5 fs and 49.5 fs are sketched in  panels (a) and (c), respectively. In panels (b) and (d) the temporal counterpart  of (a) and (c) respectively. On the bottom panels (e) and (f), detailed for each delay, the resultants pulses (red dashed line). In blue solid and green solid lines, the pulses traveling through  branches 1 and 2  \label{fig4}}
\end{figure}

\section{ Conclusions}

In conclusion,  two simple approaches we have presented, supported by experimental results, to obtain ultra-short pulses where whose FWHMs are below the Fourier limit. This technique, based on an interferometric process, shows some advantages when  compared to those commonly used for this purpose, including the capability to work with high power laser sources and operate with ultra-bandwidth pulses centered at any wavelength. The performance of the system can be summarized taking each scheme into account . For attenuation  based  scheme 7 and 12 \% of FWHM reduction were both experimentally and analytically verified, while for phase control scheme 10 and 24\%  of FWHM reduction was achieved with  main lobe to side-lobes ratios of 10 and 30\% respectively. The limitations can be associated with different devices: beam-splitter, mirrors, etc. which are affected by thermal fluctuations, mechanical vibrations, the fluence threshold of mirrors, etc., which are intrinsic properties of a Michelson interferometer. The technique,  theoretically,  works without constraints for arbitrary pulse lengths, however, experimentally it is difficult to implement the quadratic chirp parameter.  For very long pulses the dispersive media can be an optical fiber. Alternatively,  for short pulses in the few-cycle regime, other dispersive orders (third order, fourth order, etc) must be taken into account. These high orders could  be compensated with a spherical mirror \cite{reitze1992shaping}. 
We think, that this new strategy for pulse synthesis could open a new route to address ultra-fast spectroscopy, as well as a different coherent control process, from weak field to strong field domains.

 \section*{Acknowledgments}
We wish to thanks Dr. Paula Gago from the Faculty of Engineering  Imperial College UK.
E N, G T and P V are with CONICET. F V belongs to the Comisión de Investigaciones
Científicas (Buenos Aires, Argentina).
\section*{References}
\bibliographystyle{iopart-num}
\bibliography{subfourier.bib}

\end{document}